\documentstyle [aps,prd]{revtex}
\input epsf
\begin{document}

\newcommand{\alp}{$\alpha\,\,$}
\newcommand{\xe}{$x_e\;$}
\newcommand{\tdot}{$\dot{\tau}\;\,$}
\newcommand{\be}{\begin{equation}}
\newcommand{\ee}{\end{equation}}
\newcommand{\obh}{$\Omega_b h^2\;$}
\newcommand{\omh}{$\Omega_m h^2\;$}
\newcommand{\och}{$\Omega_c h^2\;$}		
\newcommand{\okh}{$\Omega_K h^2\;$}	
\newcommand{\olh}{$\Omega_\Lambda h^2\;$}	
\title{Constraining Variations in the Fine-Structure Constant with the
Cosmic Microwave Background}
\author{Manoj Kaplinghat,$^{1}$, Robert J. Scherrer,$^{1,2,3}$
and Michael S. Turner$^{3,4}$}
\address{$^1$Department of Physics, The Ohio State University,
Columbus, OH~~43210}
\address{$^2$Department of Astronomy, The Ohio State University,
Columbus, OH~~43210}
\address{$^3$NASA/Fermilab Astrophysics Center,
Fermi National Accelerator Laboratory, Batavia, IL 60510}
\address{$^4$Departments of Astronomy and Astrophysics and of Physics,
Enrico Fermi Institute, University of Chicago, Chicago, IL~~60637-1433}
\date{\today}
\maketitle

\begin{abstract}
Any time variation in the fine-structure constant alters the
ionization history of the universe and therefore changes
the pattern of cosmic microwave background fluctuations.
We calculate the changes in the spectrum of these fluctuations as a function
of the change in $\alpha$, and we find that these changes are dominated
by the change in the redshift of recombination due to the shift
in the binding energy of hydrogen.
We estimate the accuracy
with which the next generation of cosmic microwave background
experiments might constrain any variation in $\alpha$
at $z \sim 1000$.  We find that such experiments could potentially be
sensitive to
$|\Delta \alpha/\alpha| \sim 10^{-2} - 10^{-3}$.

\end{abstract}

\section{INTRODUCTION}

Physicists have long speculated that the fundamental constants
of nature are not constant, but might vary with time \cite{Dirac}.
Among the possibilities that have received the greatest attention
is the time-variation of the fine-structure constant
$\alpha \equiv e^2/\hbar c$.  The best laboratory limits
on $\Delta \alpha/\alpha$ give $|\Delta \alpha/\alpha| <
1.4 \times 10^{-14}$ over a period of 140 days \cite{prestage}.
Limits over a longer timescale can be obtained from astrophysical observations.
In particular, spectra from high-redshift quasar absorption lines
give limits of $|\Delta \alpha/\alpha| <
3 \times 10^{-6}$ at redshifts of
$z = 0.25$ and $z = 0.68$ \cite{barrow1}, and
$|\Delta \alpha /\alpha| < 3.5 \times 10^{-4}$ for $z \sim 3$ \cite{cowie},
with a claimed
detection at the level of $\Delta \alpha/\alpha = -1.5 \pm 0.3
\times 10^{-5}$ for a set of redshifts $0.5 < z < 1.6$ \cite{barrow2}.

More stringent but also more indirect limits may be placed
from geology and cosmology.  The Oklo natural nuclear reactor
yields a constraint of $-0.9 \times 10^{-7} < \Delta \alpha/\alpha <
1.2 \times 10^{-7}$, between a time of 1.8 billion years ago and
the present \cite{oklo}.  Primordial nucleosynthesis gives
$| \Delta \alpha/\alpha | < 1.0 \times 10^{-4}$ at a redshift
on the order of $10^9 - 10^{10}$ \cite{walker}.

In this paper, we consider the constraints on $\Delta \alpha/\alpha$
that could be derived from future observations of cosmic microwave
background (CMB) anisotropies.  Given the plethora of other
constraints, is there any reason to examine CMB limits
on $\Delta \alpha/\alpha$?  If $\dot \alpha$ is assumed
to be constant, then the limits quoted above correspond
to $|\dot \alpha/ \alpha| < 3.7 \times 10^{-14}$/yr (laboratory)\cite{prestage},
$|\dot \alpha /\alpha| < 5 \times 10^{-16}$/yr (quasar absorption)\cite
{barrow1}, $|\dot \alpha/\alpha| < 5-7 \times 10^{-17}$/yr (Oklo)\cite{oklo},
and $|\dot \alpha/\alpha| < 1 \times 10^{-14}$/yr (primordial
nucleosynthesis)\cite{walker}.  (Here we adopt
$H_0 = 75$ km/sec/Mpc, for consistency with ref. \cite{barrow1})
Our potential CMB limits will not be competitive with any of these.
However, in the absence of a particular model for changes
in $\alpha$, there is no reason to take $\dot \alpha$ to be constant.
Models have been proposed, for example, in which $\alpha$
oscillates \cite{marciano}.  If the value of $\alpha$
is coupled to a scalar field which evolves on cosmological
timescales, then it is conceivable that
$\alpha$ could vary as a power law in the cosmological
scale factor \cite{carroll}.  One could also imagine models in which
the scalar field evolves rapidly at early times
but later settles into a minimum, producing a fine-structure
constant which varies at high redshifts, but settles down
to a nearly constant value at low redshifts.

It is useful, therefore, to obtain limits on $\Delta \alpha/\alpha$
at redshifts $z \gg 1$.  The only limit of this type is provided
by primordial nucleosynthesis \cite{walker}; however, that
limit is very model-dependent, relying on a particular
model for the dependence of the neutron-proton mass difference
on $\alpha$.  Here we present a much more direct limit,
based on changes in the spectrum of CMB anisotropies which
could be observed by future experiments.

In the next section, we explain how changes in $\alpha$
alter the recombination scenario, and thus, the CMB
spectrum.  To simplify our discussion, we assume that $\alpha$
has a constant (different) value throughout the recombination
epoch; i.e., we neglect the possibility that $\alpha$ changes
substantially during recombination.
In Sec. 3, we calculate the $C_l$ spectrum
for different values of $\alpha$ and explain why our results
look the way they do.  In Sec. 4, we estimate the limits
which might be placed on $\Delta \alpha/\alpha$ at
$z \sim 1000$.
We find that the MAP and PLANCK experiments might be able
to reach sensitivities of $|\Delta \alpha/\alpha| \sim 10^{-2} - 10^{-3}$.

\section{Changes in the recombination scenario}\label{secrec}

The fine-structure constant $\alpha$ alters the CMB fluctuations
only to the extent that it enters into the expression
for the differential optical depth $\dot \tau$ of photons
due to Thomson scattering:
\be
\label{taudot}
\dot{\tau}=x_en_pc\sigma_T,
\ee
where $\sigma_T$ is the Thomson scattering cross-section,
$n_p$ is the number density of electrons (both free and bound)
and $x_e$ is the ionization fraction.  Thus,
$x_e n_p$ is the number density of free electrons.
The Thomson cross section depends on $\alpha$ through the relation
\be
\sigma_T=8\pi\alpha^2\hbar^2/3m_e^2c^2.\
\ee
The dependence of $x_e$ on $\alpha$ is more complicated.
Naively, one might expect $x_e$ to scale simply
with the binding energy of hydrogen, which goes as
$B = \alpha^2 m_e c^2/2$, suggesting
$x_e(T,\alpha) = x_e(T/\alpha^2)$.  We will see that
this is roughly correct, but it is not exact,
because the recombination rates depend
on $\alpha$.  The reason that $x_e$ depends on these rates
is because it does not track its equilibrium value exactly
during recombination.

Consider the standard ionization equation for Hydrogen~\cite{peebles}-
\cite{jones}:
\be -\frac{dx_e}{dt}={\cal C}\left[{\cal R}n_px_e^2-\beta(1-x_e)
\exp\left(-\frac{B_1-B_2}{kT}\right)\right], \label{ion}\ee
where ${\cal R}$ is the recombination coefficient, $\beta$ is the ionization 
coefficient, $B_n$ is the binding energy of the $n^{th}$ H-atom level and 
$n_p$ is the sum of free protons and H-atoms. The Peebles correction 
factor (${\cal C}$) accounts for the effect of
the presence of non-thermal Lyman-$\alpha$ resonance photons;
it is defined as 
\be {\cal C}(\alpha)=\frac{1+A}{1+A+C}=\frac{1+K\Lambda (1-x_e)}{1+K(\Lambda+
\beta)(1-x_e)}.\ee
In the above, \(K=H^{-1}n_pc^3/8\pi\nu_{12}^3\)
(where $\nu_{12}$ is the Lyman-\alp transition frequency) is related to the 
expansion time scale of the universe, while $\Lambda$ is the rate of decay 
of the 2s excited state to the ground state via 2 photons~\cite{spitzer}.
Clearly, $K$ scales 
as $\alpha^{-6}$ because $\nu_{12}$ scales as $\alpha^2$. Furthermore it can be ascertained that $\Lambda$ scales as 
$\alpha^8$~\cite{breit}. To investigate $\beta$, one must first use the 
principle of detailed balance to relate the ionization and recombination 
coefficients as
\be \beta={\cal R}\left(\frac{2\pi m_e kT}{h^2}\right)^{3/2}
\exp\left(-\frac{B_2}{kT}\right), \ee
while the recombination coefficient can be expressed as 
\be {\cal R}={\sum_{n, \ell}}^\star
\frac{(2\ell+1)8\pi}{c^2}\left(\frac{kT}{2\pi m_e}\right)^{3/2}
\exp\left(\frac{B_n}{kT}\right)\int_{B_n/kT}^{\infty}\frac{\sigma_{n\ell}\;y^2\,d\!y}
{\exp(y)-1}, \ee
where $\sigma_{n\ell}$ is the ionization cross-section for the 
$(n, \ell)$ excited level \cite{boardman}. 
In the above, the asterisk on the summation indicates that the 
sum from $n=2$ to $\infty$ needs to be regulated. Physically this comes about 
due to plasma effects which change the ionization and recombination 
cross-sections (calculated by considering isolated atoms). In essense, the 
summation gets truncated after a certain number of levels. For the present
purposes, it suffices to realize that the effect of this truncation scheme 
depends weakly on \alp and can be neglected~\cite{boschan}.

The \alp dependence of the cross-section ($\sigma_{n\ell}$) can be summarized 
as \(\sigma_{n\ell}\sim\alpha^{-1}f(h\nu/B_1)\), which leads to the equation:
\be \frac{\partial {\cal R}(T)}{\partial \alpha}=\frac{2}{\alpha}\left(
{\cal R}(T) - T\frac{\partial {\cal R}(T)}{\partial T}\right). 
\label{drdalp}\ee
This relation is very useful because it allows one to use the 
temperature parametrizations of ${\cal R}(T)$ in the literature. In 
particular, ${\cal R}(T)$ can be well fit by a power law of 
the form $T^{-\xi}$.
Then from equation (\ref{drdalp}), we see that the $\alpha$
dependence of ${\cal R}$ is just ${\cal R} \propto \alpha^{2(1+\xi)}$.
Let the change in \alp be 
characterized by $\Delta_\alpha \equiv \Delta \alpha/\alpha \ll 1$; 
then the corresponding fractional change in  ${\cal R}$ is 
$2\Delta_\alpha(1+\xi)$.
As it turns 
out, the results are not sensitive to the precise value of $\xi$,
which we take to be $0.7$.
Thus, to 
first order in the change in $\alpha$, it suffices to consider that 
\({\cal R}(T)\sim T^{-0.7}\).  The ionization 
equation (\ref{ion}) with the change in \alp can be expressed as
\be
\label{dxdt}
 -\frac{dx_e}{dt}={\cal C}^\prime \left[{\cal R}n_px_e^2-
\beta_{\mathrm{eff}}(1-x_e)\exp\left(-\frac{B_1-B_2}{kT}\right)\right], \ee
where \({\cal C}^\prime=(2\Delta_\alpha(1+\xi)+1)
{\cal C}(\alpha+\Delta\alpha)\) 
and $\beta_{\mathrm{eff}}$ is the effective ionization coefficient defined as
\be \beta_{\mathrm{eff}}=\beta\exp\left(-\frac{B_1}{kT}
(2\Delta_\alpha+\Delta_\alpha^2)\right).\ee

We have integrated equation (\ref{dxdt}) using CMBFAST \cite{CMBFAST}
to derive $x_e$ as as a function of redshift for several different
values of $\alpha$.  The results are displayed in Fig. 1.

\begin{figure}[Fig1]
\centering
\leavevmode\epsfxsize=12cm \epsfbox{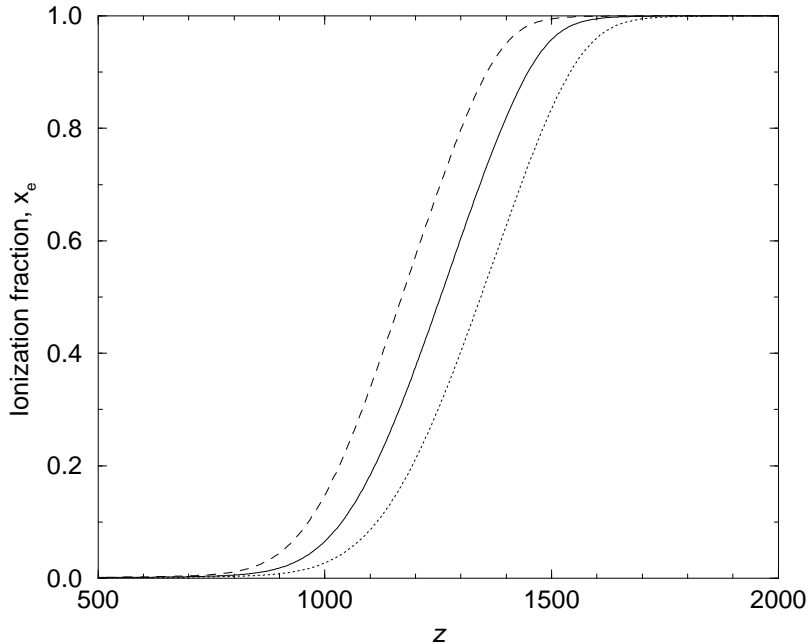}\\
\
\caption[Fig1]{\label{fig1}The ionization fraction $x_e$ as a function of
redshift $z$
for the standard scenario (SCDM, $\Omega_b = 0.05$,
$h = 0.65$) (solid curve), an increase of $\alpha$
by 3\% (dotted curve), and a decrease of $\alpha$ by 3\% (dashed curve). }
\end{figure}

The most important feature,
the shifting of $x_e(z)$ to higher $z$ when $\alpha$ is increased,
is easy to understand.  Because the equilibrium ionization fraction,
$x_e^{EQ}$, is a reasonable approximation to $x_e$,
and $x_e^{EQ} \propto (m_e/T)^{3/2}\exp (-B/T)$,
which
is dominated by the
exponential factor near recombination, to a good approximation $x_e(z)$ is simply a
function of $z/\alpha^2$ (see Fig. 2).

\begin{figure}[Fig2]
\centering
\leavevmode\epsfxsize=12cm \epsfbox{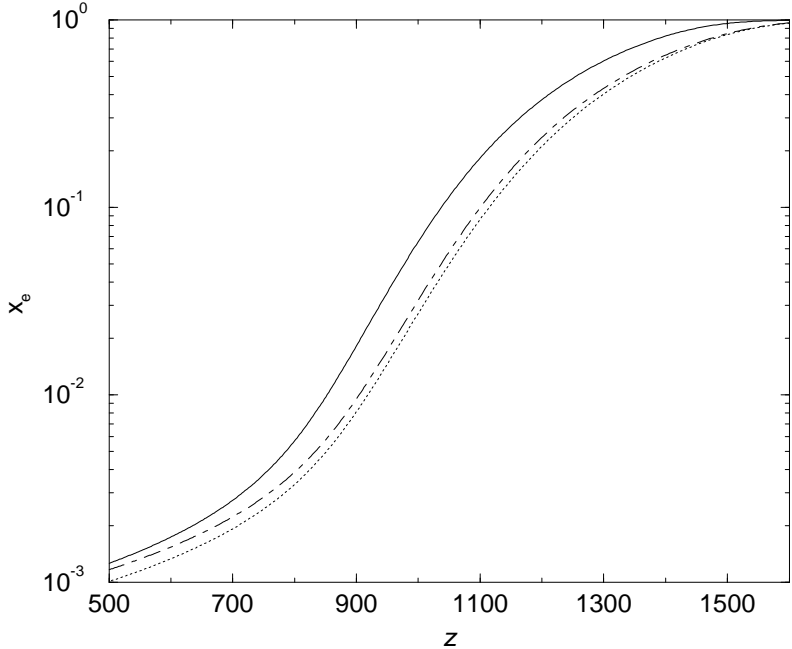}\\
\
\caption[Fig2]{\label{fig2}A comparison of the effect on $x_e$
of changing $\alpha$ by +3\% (dotted curve) with
a simple rescaling of the redshift by $\alpha^2$ (dot-dashed curve).
Solid curve is the original ionization fraction.}
\end{figure}

As can be seen in
Fig. 2, this scaling is not exact.  Two effects spoil it:
1) the factor of $(m_e/T)^{3/2}$ in $x_e^{EQ}$, and 2)
the fact that $x_e$ does not precisely track $x_e^{EQ}$.  Changing
$\alpha$ not only changes the energy levels of hydrogen, but also
all matrix elements and thereby the Thomson cross section
and recombination rates.  An increase in $\alpha$ increases the
recombination rates and so equilibrium is more closely
tracked.  (This can be seen from the fact that the residual
ionization fraction is smaller for larger $\alpha$).

More relevant for the CMB anisotropy is
the visibility function, $g(z) = e^{-\tau(z)} d\tau/dz$, which measures
the differential
probability that a photon last scattered at redshift $z$.
The visibility function depends upon $x_e$ and $\sigma_T$ through
$\tau$ (equation \ref{taudot}).  The peak of $g(z)$ defines the
location of the surface of last
scattering and its width determines the thickness of the last scattering
surface.  The finite thickness of the last scattering
surface leads to the damping of the CMB anisotropy on small
scales by smearing out temperature differences on these scales.

The shape of $g(z)$ is determined largely
by $x_e$:  around the time of last scattering, the photon mean
free path is very short until $x_e \rightarrow 0$, and the paucity of
free electrons makes Thomson scattering rare.  Increasing
$\alpha$ affects $g(z)$ in three ways:  first and most importantly,
it shifts $g(z)$ to higher redshift because $x_e^{EQ}$ is shifted
to higher redshift (by the approximate scaling
$z/\alpha^2$); second, the larger Thomson cross section increases
the opacity by an overall factor, which slightly pushes $g(z)$
to lower redshift;
and finally, the shape of the $g(z)$ curve is changed because $x_e$ more closely
tracks $x_e^{EQ}$ for larger $\alpha$.

Fig. 3 shows the visibility function expressed as a function
of conformal time, $\eta$, for different values of $\alpha$.
This is a convenient way to display it,
because the width corresponds to the comoving damping scale.  Increasing
$\alpha$ shifts $g(z)$ to higher redshift (as explained above);
at higher redshift the
expansion rate is faster, $H(z)\propto (1+z)^{3/2}$, and so
the temperature and $x_e$ decrease more rapidly, making $g(z)$
narrower.  The width of the visibility function is predicted
to scale approximately as $1/\alpha$, which is consistent with our results.

\begin{figure}[Fig3]
\centering
\leavevmode\epsfxsize=12cm \epsfbox{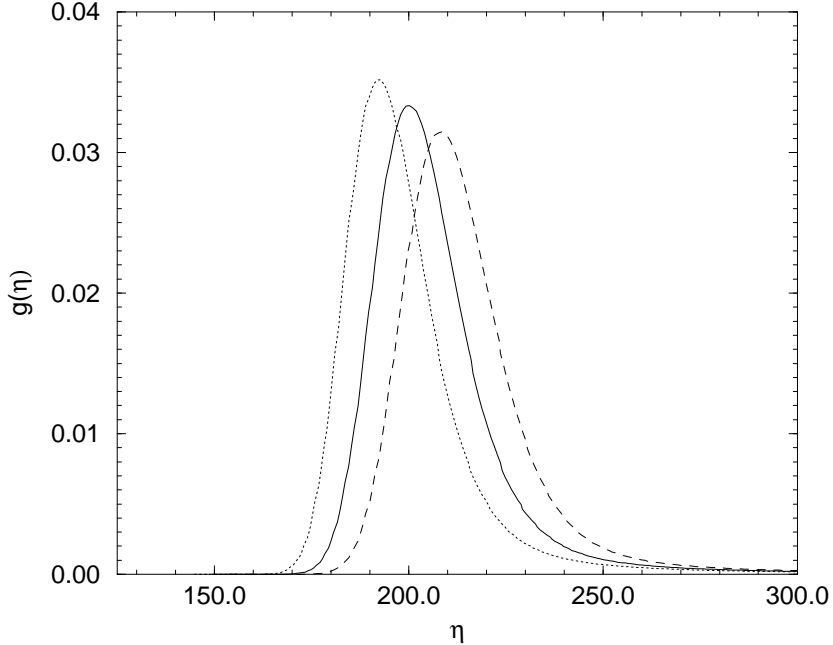}\\
\
\caption[Fig3]{\label{fig3}The visibility function $g(\eta) = e^{-\tau} d\tau/d\eta$
as a function of conformal time $\eta$ (in Mpc) for the standard scenario
(SCDM, $\Omega_b = 0.05$,
$h = 0.65$)
(solid curve), an increase of $\alpha$ by 3\% (dotted curve)
and a decrease of $\alpha$ by 3\% (dashed curve).  The peak of $g(\eta)$
defines the location of the
surface of last scattering
and its width defines the thickness of the last scattering
surface.  As can be seen, increasing $\alpha$ moves the last
scattering surface to higher redshift (smaller conformal time)
and decreases its thickness.}

\end{figure}

Are there any other potential effects on the CMB due to a variation in 
$\alpha$? One completely negligible effect is the change in the He
recombination scenario due to the change in the binding energies of He atomic 
levels. Another effect is the change in the variation of the matter 
temperature with time. Specifically, the matter temperature variation 
consists of adiabatic cooling due to the expansion of the universe and the 
cooling due to Thomson scattering. The change in $\sigma_T$ changes the 
latter. However, the matter temperature accurately tracks the radiation 
temperature until very late ($1\%$ difference at $z\sim 500$) and hence this 
effect has no consequences for the present purposes. 

\section{Changes in the CMB fluctuation spectrum}\label{secspec}

We have integrated
the changes in the differential optical depth due to a variation in \alp 
into CMBFAST \cite{CMBFAST}.
The results are shown in Fig. 4 for a $\pm 3$\% change in $\alpha$. 
Two separate effects may be noted from the results. One, for an increase in 
$\alpha$, the peak positions in the spectrum shift to higher values of 
$\ell$. Two, increasing \alp causes the values of ${\cal C}_\ell$ to 
systematically increase.  Conversely, a decrease in \alp shifts the peaks
to lower values of $l$ and decreases their amplitude.

\begin{figure}[Fig4]
\centering
\leavevmode\epsfxsize=12cm \epsfbox{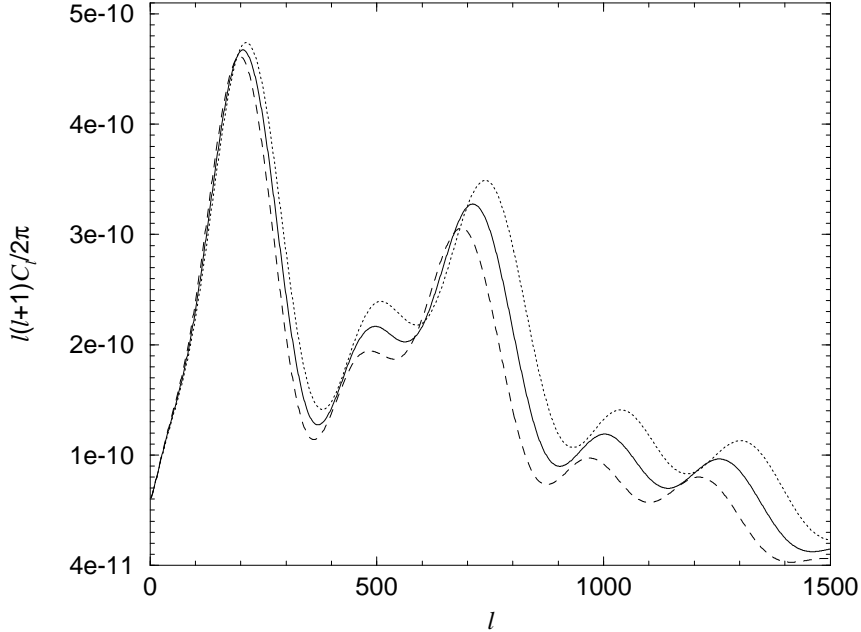}\\
\
\caption[Fig4]{\label{fig4}The spectrum of CMB fluctuations
for the standard scenario (SCDM, $\Omega_b = 0.05$,
$h = 0.65$) (solid curve), an increase of $\alpha$
by 3\% (dotted curve), and a decrease of $\alpha$ by 3\% (dashed curve) }
\end{figure}

To understand the first feature, a qualitative understanding of the position 
of the peaks is necessary. Using $\ell_p$ to denote the position of 
a peak, $r_\theta(z)$ for the angular diameter distance and $r_s(z)$ for 
the sound horizon, one can write~\cite{huthesis} 
\(\ell_p\sim r_\theta(z_{ls})/r_s(z_{ls})\), where $z_{ls}$ is the 
redshift of the surface of last scattering.
Increasing $\alpha$ increases the redshift of the last scattering
surface, as seen in Fig. 3.
A higher redshift
at the last scattering surface
corresponds to a smaller sound horizon and thus,
a higher value of $l$.  Decreasing $\alpha$ has the opposite
effect:  the redshift of last scattering decreases,
producing a larger sound horizon at last scattering, and
thus a smaller value of $l$ for the peaks.

The increase in the amplitude of the peaks with increasing
$\alpha$ derives from two separate effects.  The amplitude
of the first peak is quite sensitive to the magnitude
of the integrated Sachs-Wolfe (ISW) effect.
If a mode enters the horizon when
radiation still makes a significant contribution to the energy
density,
the decay of the gravitational
potential leads to
the blueshift of photons~\cite{hss}.  This effect
has been dubbed the ``early ISW effect" to distinguish it
from the decay of the gravitational potential at late times
in models which become dominated by curvature or a cosmological
constant.
An increase
in $\alpha$ pushes recombination to a higher redshift,
resulting
in a larger early ISW effect and, thus, a larger amplitude of the
first peak.
The early ISW effect is felt most strongly around the scale of the sound 
horizon at last scattering. For the SCDM model we have considered, this is 
around 100 Mpc or \(\ell\sim 100\). By \(\ell\sim 500\), the effect of 
early ISW contributions is negligible. 

Beyond the first peak a second effect is dominant:  diffusion
damping of CMB fluctuations due to the finite thickness of the
last scattering surface (see Fig. 3).  Because the last-scattering
surface is not infinitely thin, the anisotropies seen today
are an average over a region of finite thickness defined by
the visibility function.  This leads to damping of small-scale
anisotropies, given by a photon diffusion damping factor averaged
over the visibility function
\cite{huwhite},
\be D(\lambda)=\int_0^\infty dz\,g(z)\exp(-\lambda_D^2(z)/\lambda^2)
\approx \exp(-\lambda_D^2(z_{ls})/\lambda^2)
.\ee  
The characteristic damping length $\lambda_D$ is set by the width of the visibility
function. (The multipole damping scale is given approximately
by $l_D \sim 2H_0^{-1}/\lambda_D$).
As explained earlier and shown in Fig. 3, the comoving
damping length decreases with increasing $\alpha$.  Thus, an increase
in $\alpha$ decreases the effect of damping, and the power spectrum
at large $l$ increases with increasing $\alpha$, as seen in Fig. 4.

\section{LIMITS ON VARIATIONS IN THE FINE-STRUCTURE CONSTANT}

From the analysis presented in sections~\ref{secrec} and~\ref{secspec}, it 
is clear that a variation in \alp has a substantial effect on the CMB
fluctuation 
spectrum. The aim of this section is to obtain a quantitative measure of the 
limits put on \alp by an ideal CMB anisotropy experiment. This can be 
accomplished through an analysis of the Fisher information matrix. If our 
estimate of the cosmological parameters ($\theta_i$) is very close to the 
true values, then the likelihood function (${\cal L}$) can be expanded 
about its maximum as
\be {\cal L}\simeq{\cal L}_m\exp(-F_{ij}\delta\theta_i\delta\theta_j), \ee  
where $F_{ij}$ is the Fisher information matrix, defined as~\cite{tegmark}
\be F_{ij}=\sum_{\ell=2}^{\ell_{\mathrm max}}\frac{1}{\Delta{\cal C}_\ell^2}
\left(\frac{\partial{\cal C}_\ell}{\partial\theta_i}\right)
\left(\frac{\partial{\cal C}_\ell}{\partial\theta_j}\right). \label{fij}\ee
In equation~(\ref{fij}), the quantity $\Delta{\cal C}_\ell$ is the error in the 
measurement of ${\cal C}_\ell$.  From the Gaussian form of ${\cal L}$, the covariance matrix is seen to be 
$F^{-1}$. In particular, one can define the standard deviation for each
parameter $\theta_i$ as \(\sigma_i^2=(F^{-1})_{ii}\).
The cosmological parameters ($\theta_i$) that 
need to be determined from the measured 
fluctuation spectrum are taken to be the Hubble parameter ($h$),
the number density of baryons 
(parametrized as $\Omega_bh^2$), the cosmological constant 
(parametrized as $\Omega_\Lambda h^2$), the effective number of relativistic neutrino species 
($N_\nu$), the primordial helium mass fraction ($Y_p$), and the fine-structure constant 
($\alpha$). We make the assumption that the experiments are limited only by the 
cosmic variance up to a maximum $\ell$, denoted by $\ell_{max}$.  This assumption
is an oversimplification, but it provides a rough upper bound on the possible
limits on $\Delta \alpha/\alpha$ from future CMB experiments.

The 
fiducial models used for the present work are a standard cold dark matter 
model (SCDM) and a CDM model with a cosmological constant ($\Lambda$CDM). Both
models have $\Omega_bh^2=0.02$, $h=0.65$, $Y_p=0.246$, and $N_\nu=3.04$.
(Note that various higher-order effects, most notably the slight
heating of the $\nu \bar \nu$ pairs by electron-positron annihilation,
increase the effective value of
$N_\nu$ to 3.04 from its canonical value of 3 \cite{lopez}).
In the $\Lambda$CDM model, $\Omega_\Lambda$ is taken to be 0.7. We use
an adiabatic, scale invariant initial power spectrum and constrain the 
cosmology to be flat in keeping with the standard inflationary paradigm.
For each of these two models, we consider two limiting cases
regarding prior constraints on the unknown parameters:  first, no prior
constraints at all, and second, a ``best-case" set of limits on
the unknown parameters using priors~\cite{bond}.
In the latter case, we take, as $1-\sigma$ limits,
\(h=0.65\pm0.05\) from current observations, and 
\(\Omega_bh^2=0.02\pm0.002\) and \(Y_p=0.246\pm0.001\) from
Big-Bang nucleosynthesis \cite{ST}.  For this case, we also fixed
$N_\nu$ to be exactly equal to 3.04.

\begin{figure}[Fig5]
\centering
\leavevmode\epsfxsize=12cm \epsfbox{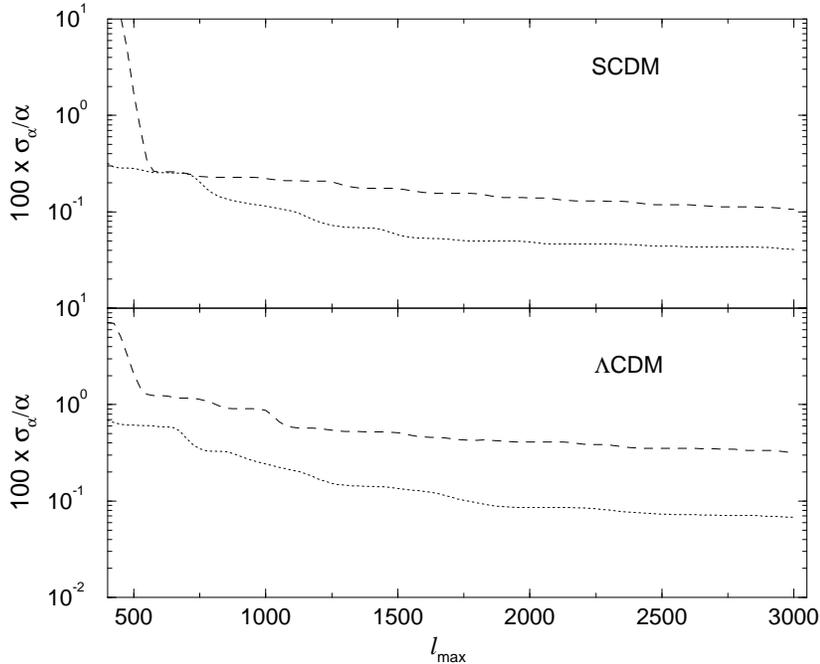}\\
\
\caption[Fig5]{\label{fig5}The estimated accuracy with which \alp can be 
constrained by a cosmic variance limited CMB anisotropy experiment, as a 
function of the maximum angular resolution given by $\ell_{max}$. The dotted
curve is the result of including priors as explained in the text, while the 
dashed curve is for the case without priors.}
\end{figure}

The required derivatives of the ${\cal C}_\ell$'s were calculated by two-sided
finite differencing for each parameter, while the rest were kept fixed. We 
verified that the changes in the results obtained were less than $10\%$ when 
the variation of the parameters was halved.
The results are shown in Fig.~\ref{fig5} in terms of the 
ratio $\sigma_\alpha/\alpha$, where $\sigma_\alpha$ is the 
$1-\sigma$ accuracy measure obtained from the Fisher matrix analysis.
We see that the estimated upper limits on $|\Delta \alpha/\alpha|$
vary from about $10^{-2}$ for $\ell_{max} \sim 500 - 1000$ down
to $10^{-3}$ for $\ell_{max} > 1500$.

These results suggest that future CMB experiments (MAP and PLANCK) might
be able to constrain any variation in the fine-structure constant
to less than $10^{-2} - 10^{-3}$.
This is a weaker constraint than can be obtained from
current quasar absorption studies, but the CMB limit would apply
at a much higher redshift ($z \sim 1000$).  It represents
a much more direct and reliable constraint than the only
other limit at $z \gg 1$, available from Big Bang
nucleosynthesis \cite{walker}.

%\acknowledgements 
\vskip 0.2in
\noindent  We are grateful to A. Heckler and S. Carroll for helpful discussions,
and we thank A. Heckler for the use of his CMB Fisher matrix code.
We thank U. Seljak and M. Zaldariagga for the use of 
CMBFAST\cite{CMBFAST}.
This work was supported in part
by NASA (NAG 5-2788) at Fermilab and
by the DOE at Fermilab, Chicago and Ohio State (DE-FG02-91ER40690).

\end{document}